\documentclass[11pt]{article}
\pdfoutput=1
\usepackage{jheppub}
\usepackage{amsmath,amssymb,bm}

\def\O{\Omega}
\def\tO{\tilde{\Omega}}
\def\tw{\tilde{w}}
\def\L{\Lambda}
\def\Ld{\ell}

\def\d{\rm{d}}
\def\rp{{r'}}

\def\nn{\nonumber}
\def\LM{{\cal{L}}_{\phi}}
\def\C{CSZ}
\def\h{I}
\def\ve{\varepsilon}
\def\p{\partial}
\def\i{{\imath}}
\def\u{\tilde{n}}

\def\tg{\tilde{g}}
\def\T{\langle\mathcal{T}\rangle} 
\def\sg{\langle\sigma\rangle}

\def\w{y}
\def\nt{{\bf{n}}}
\def\at{{\bf{a}}}

\def\K{\mathcal{K}}
\def\M{\mathcal{M}}
\def\N{\mathcal{N}}
\def\W{\mathcal{W}}
\def\R{\mathcal{R}}

\def\B{\mathcal{B}}

\def\F{\mathcal{F}}
\def\H{\mathcal{H}}

\def\DGP{\mathcal{K}}
\def\mS{\mathcal{S}}
\def\mM{{M}}

\def\Gd{\mathcal{G}^{(d+1)}}

\def\({\left(}
\def\){\right)}
\def\[{\left[}
\def\]{\right]}

\title{Emergent Dark Matter in Late Time Universe on Holographic Screen} 

\author[a,b]{Rong-Gen Cai,}
\author[c]{Sichun Sun,}
\author[d,e,f]{Yun-Long Zhang}

\affiliation[a]{CAS Key Laboratory of Theoretical Physics, Institute of Theoretical Physics, \\
Chinese Academy of Sciences, Zhong Guan Cun East Street 55, Beijing 100190, China} 

\affiliation[b]{School of Physical Sciences,
University of Chinese Academy of Sciences, \\
19(A) Yuquan Road, Shijingshan District, Beijing 100049, China}

\affiliation[c]{Center for Theoretical Sciences, Leung Center for Cosmology and Particle Astrophysics,    \\
Department of Physics,  National Taiwan University, \\
 No.1, Sec. 4, Roosevelt Road, Taipei 10617, Taiwan, China}

\affiliation[d]{Asia Pacific Center for Theoretical Physics,  APCTP Headquarters, \\
67 Cheongam-ro, Hyoja-dong, Nam-gu, Pohang 790-784, Korea}

\affiliation[e]{Center for Quantum Spacetime (CQUeST), Sogang University, \\
35 Baekbeom-ro, Sinsu-dong, Mapo-gu, Seoul 121-742, Korea}

\affiliation[f]{Yukawa Institute for Theoretical Physics, Kyoto University,\\
Kitashirakawa Oiwakecho, Sakyo-ku, Kyoto 606-8502, Japan}

\emailAdd{\!cairg@itp.ac.cn}
\emailAdd{sichunssun@gmail.com} 
\emailAdd{zhangyunlong001@gmail.com} 

\abstract{
We discuss a scenario that the dark matter in late time universe emerges as part of the holographic stress-energy tensor on the hypersurface in higher dimensional flat spacetime. Firstly we construct a toy model with a de Sitter hypersurface as the holographic screen in the flat bulk.  After adding the baryonic matter on the screen, we assume that both of the dark matter and dark energy can be described by the Brown-York stress-energy tensor. From the Hamiltonian constraint equation in the flat bulk, we find an interesting relation between the dark matter and baryonic matter's energy density parameters, by comparing with the Lambda cold dark matter parameterization. We further compare this holographic embedding of emergent dark matter with traditional braneworld scenario and present an alternative interpretation as the holographic universe. It can be reduced to our toy constraint in the late time universe, with the new parameterization of the Friedmann equation. We also comment on the possible connection with Verlinde's emergent gravity, where the dark matter is treated as the elastic response of the baryonic matter on the de Sitter spacetime background. We show that from the holographic de Sitter model with elasticity, the Tully-Fisher relation and the dark matter distribution in the galaxy scale can be derived.

}

\arxivnumber{1712.09326}

\preprint{\href{http://doi.org/10.1007/JHEP10(2018)009}{http://doi.org/10.1007/JHEP10(2018)009}}
\dedicated{\today} 
\dedicated{December 26, 2018}

\begin{document}

\maketitle


\allowdisplaybreaks

\section{Introduction}

The origin of the dark matter and the dark energy is one of the most important issues in current high energy physics and cosmology. From observations, only $5\%$ of the energy components of the current universe is visible to us. At different scales, ranging from the galactic scale to the cosmic microwave background (CMB) scale, there are many observed phenomena to test for various models of dark matter and dark energy \cite{Clifton:2011jh}. The cold dark matter model which treats the dark matter as collisionless particles is successful at CMB and large scales, but at the galactic scale, some discrepancies were proposed \cite{Famaey:2011kh, Milgrom:1983ca}. Moreover, the particle dark matter remains elusive from the direct detection so far \cite{Undagoitia:2015gya}. One of the alternatives to the cold dark matter is the modified Newtonian dynamics or modified gravity \cite{Clifton:2011jh, Famaey:2011kh, Milgrom:1983ca}, which focuses on the small scale crisis that cold dark matter cannot explain. Although those modified gravity theories seem to be less successful in producing the universe evolution scenario consistent  with CMB and large scale structure data, they can explain multiple features in galaxy rotational curves such as Tully-Fisher relation~\cite{Tully:1977fu}, Renzo's Rule~\cite{Sancisi:2003xt},  etc.

Recently, E. Verlinde proposed an emergent modified gravity  scenario from volume contribution of entanglement entropy in the de Sitter spacetime \cite{Verlinde:2016toy}, which leads to the apparent dark matter.  It is also related to the idea that  Einstein gravity can be an  emergent phenomenon  as  the entropy force with area law \cite{Verlinde:2010hp, Cai:2010hk}. Although the Verlinde's derivation in \cite{Verlinde:2016toy} received some doubts on the consistency in the literature \cite{Dai:2017qkz},  we find several Verlinde's key ideas rather inspiring. One is the possibility that our macroscopic notions of spacetime and gravity emerge from an underlying microscopic description, encouraged by the recent development of entanglement  entropy and quantum information. Another one is viewing the dark matter as merely a gravitational response of the baryonic matter on the spacetime, so as to derive the dark matter distribution around the galaxy, the Tully-Fisher relation.

In this paper, we propose a new viewpoint beyond Verlinde's emergent gravity, which can be considered as a  $(3{+}1)$ dimensional holographic screen embedded into a higher dimensional flat spacetime. We identify the holographic stress-energy tensor as that of the total dark components including the dark energy and dark matter.
We first construct a toy model, which provides a constraint relation among  the densities of dark matter, dark energy, and baryonic matter, in the case of considering the Lambda cold dark matter ($\Lambda$CDM) parameterization.
Furthermore, we generalize our toy model to the holographic Friedmann-Robertson-Walker (FRW) universe in a flat bulk and propose a new parameterization from the holographic model. The effective dark matter and dark energy are emergent and are identified with the Brown-York stress-energy tensor  \cite{Brown:1992br}. We also compare our approach to the well studied Dvali-Gabadadze-Porrati (DGP) braneworld model~\cite{Dvali:2000hr, Deffayet:2000uy, Deffayet:2001pu}.

To produce the galaxy rotational curves, we further sketch a holographic elastic model with a de Sitter boundary and fix an inconsistency in the Verlinde's paper proposed in \cite{Dai:2017qkz}. We recover the Tully-Fisher relation from the first law of thermodynamics and elasticity of the ``de Sitter medium''.
The elasticity can also be realized in blackfold approaches \cite{Emparan:2009cs, Emparan:2009at, Emparan:2009vd, Carter:2000wv, Armas:2012ac, Armas:2012jg, Armas:2014bia} or some holographic models \cite{Alberte:2015isw, Alberte:2016xja, Alberte:2017cch}. Notice here that we adopt the novel idea of elasticity of dark matter in the Verlinde's paper. Because the elasticity seems to capture the nature that the apparent dark matter is only the response of the presence of the baryonic matter. In the end, we also comment on the relation of the current construction in this paper with different scenarios such as some braneworld models and holographic models of the universe in the literature.

In section \ref{Sec2}, we firstly introduce the toy de-Sitter model in a flat bulk, which leads to the relation between dark matter component and baryonic matter component of the current universe.
In section \ref{Sec3}, we generalize our toy model to the holographic FRW universe in a flat bulk and compare it with the DGP braneworld scenario.
In section \ref{Sec4}, we reproduce the Tully-Fisher relation, with the help of holographic elasticity model and Verlinde's assumptions. We briefly compare and discuss the connection between our toy model and other scenarios, such as other braneworld models, holographic gravity, emergent gravity and summarize our results in section \ref{SecOther}.

\section{Embedding  de-Sitter Universe in a Flat Bulk} 
\label{Sec2}

We consider a $3{+}1$ dimensional time-like hypersurface with induced metric $g_{\mu\nu}$ and extrinsic curvature $\K_{\mu\nu}$, which is embedded into a $4+1$ dimensional flat bulk spacetime. 
After adding the  stress-energy tensor $T_{\mu\nu}$ of the baryonic matter and radiation, which is localized on the hypersurface,
we assume that the induced Einstein field equations on the hypersurface are modified as
\begin{align}
R_{\mu\nu}-  \frac{1}{2}R& g_{\mu\nu} - \frac{1}{L} \( \K_{\mu\nu} - \K g_{\mu\nu} \) =  \kappa_4  T_{\mu\nu}. \label{Induced1}
\end{align}
The length scale $L$ is related to the positive cosmological constant $\Lambda = {3}/{L^2}$.
The Einstein constant $\kappa_4=  {8\pi G}/{c^4}$, $G$ is the Newton gravitational constant and $c$ is the speed of light. Equivalently, we can rewrite the above modified Einstein field equations in \eqref{Induced1} as
\begin{align}
R_{\mu\nu}-  \frac{1}{2}R  g_{\mu\nu} &=  \kappa_4 T_{\mu\nu}+ \kappa_4\T_{\mu\nu} ,\label{EinsteinT1}\\
\T_{\mu\nu} &\equiv    \frac{1}{\kappa_4 L} \(  {\K_{\mu\nu}} - \K g_{\mu\nu}  \).\label{BY1}
\end{align}
They are expected to govern the late time evolution of our universe.  $\T_{\mu\nu}$ will turn out to be the Brown-York stress-energy tensor  \cite{Brown:1992br} induced from higher dimensional space time. We will see $L$ is related to the higher dimensional coupling constant $\kappa_5$ through $L=\kappa_5/\kappa_4$. At the cosmological scale, we assume that $T_{\mu\nu}$ only includes the stress-energy tensor of baryonic matter and radiation. While $\T_{\mu\nu}$ in \eqref{BY1} represents the total dark components in our universe, such as the dark energy and dark matter.

We are going to  consider the parameterization in $\Lambda$CDM model describing the evolution of the  late time universe. In detail, we take the FRW metric in $3+1$ dimensions, which assumes that our universe is uniform and isotropic at large scale, with scale factor $a(t)$, 
\begin{align}\label{FRW1}
{\d}s_{4}^2 &=g_{\mu\nu}{\d}x^\mu{\d} x^\nu =  -c^2 {\d}t^2 +a(t)^2\[\frac{{\d}r^2}{1- k r^2}  +r^2 {\d} \O_{2}\].
\end{align}
In the spatial flat $\Lambda$CDM model with $k=0$, it contains a positive cosmological constant $\Lambda$, which contributes to the dark energy with density parameter $\Omega_{\L}$, cold dark matter density parameter  $\Omega_D$, and baryon density parameter $\Omega_B$.
The Friedmann equation is given by
\begin{align} \label{FriedmannR}
 \frac{H(t)^2}{H_0^2} =  \Omega_\Lambda +  \frac{\Omega_D}{a(t)^3} + \frac{\Omega_B}{a(t)^3} + \frac{\Omega_R}{a(t)^4}.
\end{align}
$H(t)$ is the Hubble parameter and $H_0\equiv H(t_0)$ is the Hubble constant today at $t=t_0$.
 \begin{align}  \label{rhoC}
 H(t)\equiv  \frac{\dot{a}(t)}{a(t)}, \qquad  H_0^2  = \frac{\kappa_4 c^4 }{3}  \rho_{c} ~\Rightarrow~ \rho_{c} =\frac{3 }{ \kappa_4}\frac{H_0^2}{c^4}.
\end{align}
$\dot{a}(t)$ is the derivative with respect to the time $t$ and $\rho_{c} $ is the critical energy density of the universe.
If requiring $a(t_0)=1$, from \eqref{FriedmannR} we have  $1=\Omega_{\L}  +\Omega_B +\Omega_D +\O_{R} $.
Since the radiation density parameter $\Omega_R\sim 10^{-4}$ is very small,
in the late time universe we will simply take $\Omega_D+\Omega_B+\Omega_{\L}\simeq 1$.
After neglecting the radiation component $\O_R$ in \eqref{FriedmannR}, the Friedmann equation of the late time spatial flat $\Lambda$CDM  universe can be written as
 \begin{align}\label{LCDMF}
 \frac{H(t)^2}{H_0^2} =  \Omega_\Lambda +  \frac{\Omega_D}{a(t)^3} + \frac{\Omega_B}{a(t)^3}
 \equiv \tO_\L +\tO_D+\tO_B.
 \end{align}
We have introduced the time dependent notations with tilde, which satisify
 \begin{align}\label{omega1}
\Omega_{\L}=\tO_\L |_{t=t_0},\quad \Omega_{D}=\tO_D |_{t=t_0}, \quad \Omega_{B}=\tO_B |_{t=t_0} .
\end{align}

Based on the modified Einstein field equations \eqref{Induced1} and the Hamiltonian constraint from the consistent embedding in higher dimensional flat bulk, we are going to show an interesting constraint relation between these parameters,
\begin{align}
\text{\C:} \quad\Omega_{D}^2&=\frac{1}{2} \Omega_{\L} (\Omega_{D}- \Omega_B). \label{CSZ1}
\end{align}
Let us compare with the constraint relation in the Verlinde's emergent gravity \cite{Verlinde:2016toy},
\begin{align}
 \text{Verlinde:}\quad  \O_D^2& = {4\over 3}  \O_B. \label{Verlinde1}
\end{align}
In the current universe both of these two relations \eqref{CSZ1} and \eqref{Verlinde1} are remarkably well obeyed. 
Taking the observation values within the $\Lambda$CDM model~\cite{Ade:2015xua, Abbott:2017wau}, with a bit priori choice of the parameters as
 \begin{align}\label{OmegaD}
\Omega_{\L}\simeq0.685,\quad \Omega_D\simeq0.265,  \quad \Omega_B\simeq0.050 ,
\end{align}
we can calculate the following differences,  
\begin{align}
  \delta_{\C}&\equiv\Omega_D^2 - \frac{1}{2} \Omega_{\L} (\Omega_D- \Omega_B)\simeq-0.003  \, ,  \label{DeltaCSZ}\\
 \delta_V&\equiv\Omega_D^2 - {4\over 3} \Omega_B \simeq 0.004 \, . \label{DeltaV} 
 \end{align}
Thus, our relation \eqref{CSZ1} holds as well as the Verlinde's \eqref{Verlinde1} with minor difference in approximation.
We will show  how to derive this constraint equation \eqref{CSZ1} in the following sections.

\subsection{Hamiltonian Constraint From Hypersurface Embedding}

Similar to the formula \eqref{EinsteinT1},
let us write down the Einstein equation in $d$ dimensional spacetime as
\begin{align}
R_{\mu\nu}-  \frac{1}{2} R\,  g_{\mu\nu}=\kappa_{d} \[ T_{\mu\nu} + \T_{\mu\nu} \],\label{Einsteind}
\end{align}
with $\mu,\nu=0,1,...,(d-1)$, and $\kappa_{d}=  {8\pi G_{d}}/{c^4}$.
$T_{\mu\nu}$ is the stress-energy tensor of baryonic matter and radiation,
and $\T_{\mu\nu}$ is the effective dark components of our universe, which can include both of the dark energy and dark matter.
The trace of above equations yields the Ricci scalar
\begin{align}
R & = -\frac{2\kappa_{d} }{d-2} \[ T +\T \].\label{RicciS1}
\end{align}

Now we assume that the geometry with metric $g_{\mu\nu}$ can be embedded into one higher dimensional spacetime, as a hypersurface with the normal vector $\N^{A}$, and the indices $A, B = 0,1,...,d$.
We can define the induced metric on the hypersurface $g^A_{~~B}= \tg^A_{~~B}-\N^{A}\N_{B}$
as well as the extrinsic curvature  ${\K_{\mu\nu}} \equiv g_\mu^{~A} g_\nu^{~B} \tilde{\nabla}_{(A} \N_{B)}$,
 with $\mu, \nu$ are the indices on the hypersurface, which depend on the coordinate choices.
 $\tilde{\nabla}$ is the covariant derivative associated with the bulk metric $\tg_{AB}$.
Even though there exists  matter in the late time universe, we require them to be localized on the hypersurface,
such that we still have $\Gd_{AB} \N^{A} \N^{B}=\mathcal{T}_{AB}^{(d+1)} \N^{A} \N^{B}=0$ .
Thus, considering the Gauss equations, the Hamiltonian constraint equation of the hypersurface leads to
\begin{align}\label{Hamiltonian1}
0= 2\,  \Gd_{AB} \N^{A} \N^{B}   &\equiv \K^2- {\K_{\mu\nu}}\K^{\mu\nu}  -R.
\end{align}
On the other hand, the momentum constraint equations  $\Gd_{AB} \N^{A} {g^N}_{\nu}=\mathcal{T}_{AB}^{(d+1)} \N^{A} {g^N}_{\nu}=0$ lead to
\begin{align}\label{momentum1}
0=\Gd_{AB} \N^{A} {g^N}_{\nu}&= \nabla_\mu \( {\K^\mu}_\nu - \K {g^\mu}_\nu \)=0.
\end{align}
$\nabla$ is the covariant derivative associated with the metric $g_{\mu\nu}$ on the hypersurface.

Next we assume that the stress-energy tensor of the dark components in \eqref{Einsteind} can be given by the Brown-York stress-energy tensor  associated with the hypersurface~\cite{Brown:1992br},
\begin{align}
\T_{\mu\nu}& =\frac{1}{\kappa_{d+1}} \( {\K_{\mu\nu}} - \K g_{\mu\nu} \),\label{BYtensor}
\end{align}
and $\kappa_{d+1}$ is the Einstein's constant in $d+1$ dimensions.
Replacing the extrinsic curvature by the Brown-York stress-energy tensor,
the Hamiltonian constraint relation \eqref{Hamiltonian1} becomes
\begin{align}
 \frac{\T^2}{d-1}-\T_{\mu\nu}\T^{\mu\nu} = \frac{ R   }{( \kappa_{d+1})^2}\, .
 \label{Constraint1}
\end{align}
Then by plugging \eqref{RicciS1} into \eqref{Constraint1}, we have
\begin{align}
\frac{\T^2}{d-1}-\T_{\mu\nu}\T^{\mu\nu} =
- \frac{\kappa_d}{( \kappa_{d+1})^2} \frac{2}{d-2} \(T+ \T \).
\label{Constraint0}
\end{align}
While the momentum constraint equations \eqref{momentum1} lead to $\nabla^\mu \T_{\mu\nu}=0$.

\subsection{Holographic de-Sitter Screen in a Flat Bulk}  

Firstly we set that the stress-energy tensor of the baryonic matter and radiation in the Einstein field equations \eqref{Einsteind} vanish, $T_{\mu\nu}{=}0$.
As a warm up, let us consider the hypersurface as the $d$ dimensional de Sitter spacetime,
\begin{align}
{\d} s_{d}^2 &=g_{\mu\nu}{\d}x^\mu{\d} x^\nu= -c^2 {\d}t^2 +e^{2(ct/L)}\[ {{\d}r^2}  +r^2 {\d}\O_{d-2}\],
\label{dS0}
 \end{align}
which can be embedded into the $d+1$ dimensional flat spacetime
  \begin{align}
{\d} s^2_{d+1} &
=\eta_{AB}{\d} x^{A}{\d} x^{B}=- {\d} X_0^2+{\d} X_i^2,\label{flat0}
 \end{align}
with $i=1,2,...,d$.
The vacuum Einstein field equations associated with the $(d{+}1)$-dimensional flat bulk metric \eqref{flat0}
turn out to be $\Gd_{AB}=0$.
Let us study the embedding of de Sitter hypersurface \eqref{dS0} in more details.
It is a hyperbolic spacetime with radius $L$,
\begin{align}
L^2=- T^2+ X_i^2,\qquad \N^{A}=\frac{1}{L}(X_0,X_i).\label{flat1}
\end{align}
where $\N^{A}$ is the normal vector of the hypersurface pointing outwards.
The cosmological constant $\Lambda_d =\frac{(d-1)(d-2)}{2 L^2}$ will play the role of the dark energy.
Notice that to balance the Einstein field equations \eqref{Einsteind} with the induced de Sitter metric $g_{\mu\nu}$ in \eqref{dS0}, one requires either the cosmological constant or the apparent dark energy term.

Interestingly, for the pure de Sitter spacetime \eqref{dS0},
after considering  normal vector \eqref{flat1} which leads to the extrinsic curvature $ {\K}^{\mu\nu}=\frac{1}{L} g^{\mu\nu}$,
the Brown-York stress-energy tensor  \eqref{BYtensor} turns out to be $\T^{\mu\nu} = \T^{\mu\nu}_{\L} = - \frac{1}{\kappa_{d+1}} \frac{d-1}{L} g^{\mu\nu}$.
Then we arrive at the stress-energy tensor of apparent dark energy,
 \begin{align}\label{kappaD}
 \T^{\mu\nu}_{\L}&= - \frac{\Lambda_d}{\kappa_d} g^{\mu\nu},\quad
 \text{when} ~\frac{  \kappa_{d+1}}{ \kappa_{d}}  = \frac{2 L}{d-2} \,.
\end{align}
One can see that the Einstein field equations in \eqref{Einsteind} with the de Sitter metric \eqref{dS0} are naturally satisfied
\begin{align}
R_{\mu\nu}-  \frac{1}{2} R\,  g_{\mu\nu}=\kappa_{d}   \T_{\mu\nu}^{\L} .
\end{align}

From \eqref{kappaD} we read out the dark energy density formula
 \begin{align}\label{kappar}
\tilde\rho_{\L}  = \frac{u_\mu u_\nu }{c^4} \T^{\mu\nu}_{\L} = \frac{\Lambda_d}{\kappa_d c^2}\, .
 \end{align}
After considering \eqref{Constraint0} with $T=0$, we have the identity
\begin{align}   \label{Background}
\frac{{\T}_{\!\L}^2}{d-1}-{\T}^{\!\L}_{\mu \nu}{\T}_{\!\L}^{\mu\nu}  = - \frac{\tilde\rho_{\L} c^2}{d-1}  \T_{\!\L} .
\end{align}
Thus, assuming $\T^{\mu\nu}  =\T^{\mu\nu}_{\L}\equiv - \frac{\Lambda}{\kappa_d} g_{\mu\nu}$ in the constraint equation \eqref{Constraint1},
 the pure de Sitter spacetime satisfies the above identity automatically. Notice here that  the Brown-York stress-energy tensor  plays the role of dark energy and there is no baryonic matter or dark matter yet in the set-up.

\subsection{Emergent Dark Matter on Holographic Screen}

Next, we consider to add a small amount of baryonic matter and radiation in with the uniform and isotropic distribution.
Considering \eqref{kappaD}\eqref{kappar},
our assumption for the constraint relation \eqref{Constraint0} becomes
\begin{align} \frac{\T^2}{d-1}-\T_{\mu\nu}\T^{\mu\nu} = - \frac{\tilde\rho_{\L} c^2}{d-1} \big[ T+ \T\big].
\label{Constraint2}
\end{align}
This is the main constraint relation in this section.
Since in Einstein field equations \eqref{Einsteind},
$T_{\mu\nu}$ is the stress-energy tensors of baryonic matter and radiation,
\begin{align}
T^{\mu\nu}  &=T_B^{\mu\nu} +T_R^{\mu\nu},\qquad T_B^{\mu\nu}  =(\rho_B  )u^\mu u^\nu,\quad T_R^{\mu\nu}  =(\rho_R  )u^\mu u^\nu+ p_R h^{\mu\nu},
\end{align}
where $\rho_B$ is the mass density baryonic matter, $h^{\mu\nu}=g^{\mu\nu} +u^\mu u^\nu$ and $u^\mu$ is the velocity in $d$ dimensions. The dark energy and dark matter are all assumed to be related to the extrinsic curvature of the hypersurface embedded in the higher dimensional flat bulk.
We take the Brown-York stress-energy tensor $\T_{\mu\nu}$, which is playing the role of dark energy and dark matter, 
\begin{align}\label{BYDM}
\T^{\mu\nu} &=  \T_{\!\L}^{\mu\nu}+\T_D^{\mu\nu},\qquad  
\T_{\!\L}^{\mu\nu}= -(\rho_{\L} c^2)  g^{\mu\nu},\quad \T_D^{\mu\nu}=(\rho_D ) u^\mu u^\nu+p_D h^{\mu\nu}.
\end{align}
Putting them back into the constraint equation \eqref{Constraint2}, we have
 \begin{align}
&(\rho_\Lambda+\rho_D) \Big[ d\rho_\Lambda -(d-2) \rho_D- 2(d-1)  \frac{p_D}{c^2} \Big] \nn \\
=~& \tilde\rho_{\L} \Big\{ d\rho_{\L}+\rho_B+\Big[ \rho_D - (d-1)\frac{p_D}{c^2}\Big]+\Big[ \rho_R - (d-1)\frac{p_R}{c^2}\Big] \Big\}.
\end{align} 
If setting $ \tilde\rho_{\L}=\rho_{\L}$  and with equation \eqref{Background}, we arrive at
\begin{align}
 \rho_D^2& = \frac{\rho_{\L}}{d-2}  \[\rho_D-\rho_B-\rho_R+(d-1)\frac{p_R}{c^2}  \]
 -\frac{d-1}{d-2}\frac{p_D}{c^2}\(2\rho_D+ \rho_\Lambda \). \label{CSZ2}
\end{align}

When $d=4$, the stress-energy tensor of radiation is traceless  $-\rho_R c^2+  3 p_R =0$.
Keeping the pressure $p_D$ of the dark matter in the constraint relation \eqref{CSZ2} leads to
 \begin{align}
 \rho_D^2& = \frac{\rho_{\L}}{2 (1+3 {\tw}_{D})}  \big[ \rho_D(1-3{\tw}_{D})-\rho_B \big],
 \qquad {\tw}_{D} \equiv \frac{p_D}{\rho_D c^2} \, . \label{CSZ3}
\end{align}
${\tw}_{D}$ denotes the effective state equation of the emergent dark matter, which can be time dependent in general.
Dividing both sides of \eqref{CSZ3} by the squire of the critical energy density $\rho_c^2$ in \eqref{rhoC}, we obtain the generalized constraint relation
 \begin{align}\label{CSZ5}
\tO_{D}^2 =\frac{ \tO_{\L}}{2 (1+3 {\tw}_{D}) } \big[  \tO_{D}(1-3 {\tw}_{D}) - \tO_B\big].
 \end{align}
The components have been identified as
 \begin{align}\label{Omega1}
\tO_{\L}\equiv  {\rho_\L}/{\rho_c},\quad
\tO_{D}\equiv  {\rho_D}/{\rho_c},\quad
\tO_{B}\equiv  {\rho_B}/{\rho_c},
 \end{align}
 which can be  time dependent in general case.

We will take the assumption that the evolution of the late time universe is governed by the $\Lambda$CDM parameterization,
and the total dark components are identified as the Brown-York stress-energy tensor in \eqref{BY1}.
We also assume  the emergent  dark matter is pressureless at $t=t_0$ for now and discuss the otherwise later in this paper.
Through setting ${\tw}_{D}= 0$ in \eqref{CSZ5}, and considering \eqref{omega1}, we can obtain our main toy constraint in \eqref{CSZ1},
 \begin{align}\label{CSZ4}
\Omega_{D}^2 =\frac{1}{2} \Omega_{\L} (\Omega_{D}- \Omega_B).
 \end{align}

If we further use $\Omega_{\L} +\Omega_D +\Omega_B\simeq 1$ in the late time universe, then
\begin{align}
\Omega_{D}^2 =&\frac{1}{3}  (\Omega_{D}- \Omega_B+ \Omega_B^2).
 \end{align}
Considering $\Omega_{D}\simeq 5 \Omega_B$ from \eqref{OmegaD}, as well as $\Omega_B\simeq0.05 \ll 1$,
 we can also arrive at the Verlinde's  $\Omega_{D}^2 \simeq\frac{4}{3} \Omega_B$ in \eqref{Verlinde1}.
On the other hand, since $\Omega_B+ \Omega_{D} \lesssim\Omega_{\L}$,
despite being not so precise, the de Sitter background is still a good approximation. However, if we consider the dark matter in smaller scales around the galaxies and compare with galactic rotational curves, we need to consider the effects of back-reaction of baryonic matter. This is the same situation in the earlier universe when matter or radiation dominates  in 
the energy of the universe and can not be treated as perturbations on the background anymore. In such cases, this toy model turns out to be not enough, we will resort to the more complicated model in the next section.

\section{Holographic FRW Universe and Emergent Dark Matter}
\label{Sec3}
In this section, we consider a more consistent embedding of the FRW metric into one higher dimensional flat spacetime \cite{Maartens:2010ar}. We assume that the stress-energy tensor of the total dark components, including dark matter and dark energy, is provided by the holographic stress-energy tensor on the FRW hypersurface.
In  section \ref{sec31}, we firstly review the consistent embedding of the FRW hypersurface in a flat bulk. In section \ref{sec32}, we review the usual parameterization in DGP braneworld model with the $Z_2$ symmetry along the hypersurface. In section \ref{sec33}, we further develop our viewpoint on the holographic FRW universe in a flat bulk, using a different boundary condition from \ref{sec32}. We also discuss its connection to the well studied DGP braneworld model. In particular, we show that under  special parameter choice, the constraint relation \eqref{CSZ1} in our toy model can be recovered in late time universe.

\subsection{Embedding FRW Universe in a Flat Bulk}
\label{sec31}

Consider a $4+1$ dimensional flat bulk $\M$ with action ${\mS}_{5}$ and metric $\tg_{AB}$, along with the  $3+1$ dimensional time like hypersurface $\p\M$ with action ${\mS}_{4}$ and induced metric $g_{\mu\nu}$.
The total action is given by ${\mS}_{5}+{\mS}_{4}$, where
\begin{align}\label{DGP}
&{\mS}_{5}=\frac{1}{2 \kappa_{5}}\int_{\M} {\d}^{5} x \sqrt{-\tilde{g}} \, \R  + \frac{1}{\kappa_{5}}\int_{\p\M} {\d}^{4}x \sqrt{-{g}}\,\K, \\
&{\mS}_{4}= \frac{1}{2 \kappa_4}\int_{\p\M} {\d}^4x \sqrt{-{g}} \,R + \int_{\p\M} {\d}^4x \sqrt{-{g}}{\cal{L}}_{\mM}  \, .
\end{align}
$\K$ is the trace of extrinsic curvature of the hypersurface $\p\M$, and ${\cal{L}}_{\mM}$ is the Lagrange density of matter localized on the hypersurface.
If choose the Gaussian normal coordinates of the bulk metric $\tg_{AB}$, we have
\begin{align}\label{Gaussian}
{\d} s^2_5 =\tg_{AB} {\d}x^{A} {\d}x^{B} =  {\d} \w^2 + \tg_{\mu\nu} {\d}x^\mu {\d}x^\nu .
\end{align}
We assume the hypersurface $\p\M$ located at $\w=0$, which is the shared boundary of the half bulk $\M_+$ for the region $\w>0$ and the half bulk $\M_-$ for the region $\w<0$.

The bulk equations of motion are given by the variation of the total action ${\mS}_{5}+{\mS}_{4}$ with the bulk metric $\tg_{AB}$,
\begin{align}
\frac{1}{\kappa_5}\(\R_{AB}-\frac{1}{2}\R \tilde{g}_{AB}\) +\frac{1}{\kappa_4}\( R_{\mu\nu}-\frac{1}{2} R g_{\mu\nu}\){{\tg}^\mu}_{~A}{{\tg}^\nu}_{~B} \delta{(\w)} &=  T_{\mu\nu}^{\mM} {\tg^\mu}_{~A}{\tg^\nu}_{~B} \delta{(\w)} \,.
\end{align}
With the matching junction condition at the hypersurface $\w=0$,
\begin{align}\label{junction1}
 \T_{\mu\nu}^{+} - \T_{\mu\nu}^{-}  + \frac{1}{\kappa_4} G_{\mu\nu}&=T_{\mu\nu}^{\mM},
\end{align}
where $ G_{\mu\nu}\equiv R_{\mu\nu}-\frac{1}{2} R g_{\mu\nu} $.
The effective stress-energy tensor from extrinsic curvature is
\begin{align} \label{Junction}
\T_{\mu\nu}^{\pm} \equiv \frac{1}{\kappa_5}  \( {\K^\pm_{\mu\nu}} -\K^\pm g_{\mu\nu}  \).
\end{align} 
We include the baryonic matter, radiation and other effective terms in the Lagrangian ${\cal{L}}_{\mM}$, which leads to the stress-energy tensor
\begin{align}
T_{\mu\nu}^{\mM} &=-\frac{2}{\sqrt{-g}} \frac{ \delta }{\delta g^{\mu\nu}}\( \int_{\p\M} d^4x \sqrt{-{g}}{\cal{L}}_{\mM}\).
\end{align}
The extrinsic curvature is $\K^{\pm}_{\mu\nu}\equiv \tg_\mu^A \tg_\nu^B \tilde{\nabla}_{(A} \N^{\pm}_{B)}  |_{\p\M}$, and $\N^\pm$ is chosen as the normal vector of $\p\M$ along the $\pm\w$ directions, respectively.

We consider that our universe is uniform and isotropic at large scale, and take the spatially flat FRW metric in $d=4$ dimensions, 
   \begin{align}\label{FRW2}
{\d} s^2_4 =& -c^2 {\d} t^2 +a(t)^2\[{\d}{r}^2 +{ r}^2 {\d}\O_{2}\].
\end{align}
The consistent embedding in higher dimensional flat  spacetime has been discussed in \cite{Binetruy:1999ut},
where the bulk metric \eqref{Gaussian} in Gaussian normal coordinates is
\begin{align}\label{Gaussmetric1}
{\d} s^2_5 =\tg_{AB} {\d}x^{A} {\d}x^{B} =  {\d} \w^2  - \nt(\w,t)^2 \  c^2 {\d} t^2 +\at(\w,t)^2\[{\d}{r}^2 +{ r}^2 {\d}\O_{2}\] .
\end{align}
The consistent embedding metric functions are solved as  \cite{Dick:2001sc, Dick:2001np, Lue:2002fe},
\begin{align}
\at(\w,t)^2 &=a(t)^2 +  \w^2  \frac{\dot{a}(t)^2}{c^2} \pm 2 \w \sqrt{a(t)^2\frac{ \dot{a}(t)^2}{c^2}+{\h} } ,\label{Gat}\\
\nt(\w,t)  &=\frac{\p_t{\at}(\w,t)}{\dot{a}(t)} \, .\label{Gnt}
\end{align}
Here ${\h}$ is the integration constant, with dimension of $[L]^{-2}$.

In the following section \ref{sec32}, we will choose the $Z_2$ symmetry along the brane similar as the usual DGP model, where the parameter ${\h}$ is neglected.
 In section \ref{sec33}, we will consider the FRW brane as a cutoff hypersurface in the flat bulk and present an alternative interpretation  as the holographic FRW(hFRW) universe, with a non-zero parameter ${\h}$.

\subsection{Braneworld Scenario: DGP Model with $Z_2$ Symmetry}
\label{sec32}
 
In the usual DGP model~\cite{Dvali:2000hr, Deffayet:2000uy, Deffayet:2001pu},  the $Z_2$ symmetry along the brane in the bulk has been imposed,
which leads to the boundary condition $\T^{ +}_{\mu\nu}=-\T^{ -}_{\mu\nu}$, as well as the modified Einstein field equations on the brane
\begin{align}
 \frac{1}{\kappa_4} G_{\mu\nu}&=T_{\mu\nu}^{\mM} + \T_{\mu\nu}^{{\DGP}}\,,\quad
  \T_{\mu\nu}^{{\DGP}}  \equiv   \T^{ -}_{\mu\nu}- \T^{ +}_{\mu\nu}= \frac{2}{\kappa_5} \({\K^-_{\mu\nu}} -\K^- g_{\mu\nu}\).\label{TmnH}
\end{align}
In the self accelerating branch of the DGP model (sDGP),
$T_{\mu\nu}^{\mM}$ includes the baryonic matter and dark matter, while $\T_{\mu\nu}^{\K}$ provides the effective dark energy.
In the normal branch of the DGP model (nDGP),
depending on the parameterization, a cosmological constant needs to be supplemented.
In the usual setup, ${\h}=0$ in \eqref{Gat} was chosen, then the metric \eqref{Gaussmetric1} becomes more simple,
\begin{align}\label{Gaussmetric2}
{\d} s^2_5  &= {\d} \w^2 - \[1\pm \frac{|\w|}{c} \frac{\ddot{a}(t)}{\dot{a}(t)} \]^2 c^2 {\d} t^2 +\[1\pm \frac{ |\w|}{c} \frac{\dot{a}(t)}{a(t)} \]^2 a(t)^2\({\d}{r}^2 +{r}^2 {\d}\O_{2}\).
\end{align}
The above two equations will lead to the modified Friedmann equation on the brane,
\begin{align}\label{Friedmann1}
H(t)^2 &= \frac{\kappa_4 c^4 }{3} \[ \rho_{\mM} (t) + \rho_{\DGP}(t)\],
\end{align}
as well as the acceleration equation
\begin{align}\label{acceleration1}
\dot H(t)+H(t)^2&=- \frac{\kappa_4 c^4 }{6}  \[ \rho_{\mM} (t) + \rho_{\DGP}(t)+ 3 \frac{ p_{\mM} (t) + p_{\DGP}(t)}{c^2}\].\end{align}
The energy conservation equations for each component are
\begin{align}\label{conservation1}
\dot\rho_{\i}(t)&=-3 H(t) \[\rho_\i(t)+ p_\i(t)/c^2\],\qquad \i=\mM, \, {\DGP}\,.
\end{align}
Plugging the metric \eqref{Gaussmetric2} into the stress-energy tensor $\T_{\mu\nu}^{{\DGP}}$ in \eqref{TmnH}, we can read out the effective energy density and pressure
\begin{align}\label{rhop1}
\rho_{\DGP}(t)& =\pm\frac{2}{\kappa_5 c^3}  3H(t), \\
p_{\DGP}(t) &=\mp \frac{2}{\kappa_5 c }  \Big[ 3 H(t)  + \frac{\dot{H}(t)}{H(t)}\Big].
\end{align}

The positive $\rho_{\DGP}$ and negative $p_{\DGP}$ correspond to the sDGP branch. The negative $\rho_{\DGP}$ and positive $p_{\DGP}$ correspond the nDGP branch, and the extra effective cosmological constant is required in equation \eqref{Friedmann1}.
Since $ \T_{\mu\nu}^{{\DGP}}$ in \eqref{TmnH} is proportional to the Brown-York stress-energy tensor,
it is natural to see that the Hamiltonian constraint equation on the brane in the bulk \eqref{Hamiltonian1} is satisfied and leads to
\begin{align}
\frac{\T_{\DGP}^2}{3}-\T^{\DGP}_{\mu\nu}\T_{\DGP}^{\mu\nu} =
- \frac{ \kappa_4}{( \kappa_{5}/2)^2} \[ T_{\mM} + \T_{\DGP} \].
\label{ConstraintDGP}
\end{align}
Compared with the constraint relation in our toy model \eqref{Constraint0}, the coupling constant $\kappa_{5}$ is replaced by $\kappa_{5}/2$ in \eqref{ConstraintDGP}. It is due to the double copies of the Brown-York stress-energy tensor in $\T^{\DGP}_{\mu\nu}$  at \eqref{TmnH} in the DGP model.

Here we pay attention to the sDGP branch, where the modified Friedmann equation \eqref{Friedmann1} is summarized as
\begin{align}  \label{DGPF1}
 \frac{H(t)^2}{H_0^2} = \frac{\Omega_M}{a(t)^3}  + \sqrt{\Omega_{\Ld}} \frac{H(t)}{H_0},
\qquad  \Omega_{\Ld}=\frac{c^2}{\Ld^2 H_0^2},\qquad \Ld\equiv\frac{(\kappa_5/2)}{\kappa_4}.
\end{align}
 If considering $a(t_0)=1$ and $H(t_0)=H_0$ in \eqref{DGPF1},
we have $1=\Omega_M+ \sqrt{\Omega_{\Ld}}$.
Compared with \eqref{kappaD} in our toy model, $\Ld=L/2$ is due to the $Z_2$ symmetry of the sDGP brane, which includes the double copies of the Brown-York stress-energy tensor.
Equivalently,
\begin{align} \label{DGPF}
\frac{H(t)^2}{H_0^2}   =  \frac{\Omega_{\Ld}}{2}+ \frac{\Omega_M}{a(t)^3}
+  \[ \frac{\Omega_{\Ld}^2 }{4}+ \frac{\Omega_{\Ld} \Omega_M}{a(t)^3}   \]^{1/2},\qquad
\O_M=1-\sqrt{\O_\Ld} .
\end{align}
Notice that to make the presentation more clear, we did not include the contribution from spatial curvature $\Omega_K$ and radiation  ${\Omega_R}$.
It is easy to see that if setting $\Omega_M = 0$ in \eqref{DGPF}, we will have the Friedmann equation of the de-Sitter Universe.
In the self accelerating branch of the DGP model, $\Omega_M=\Omega_B + \Omega_D$, including both of the components of baryonic matter and dark matter, while $\Omega_\ell$ is the component of the effective dark energy.
More detailed study of the phenomenology of the DGP model can be found in \cite{Lue:2005ya, Song:2006jk, Fang:2008kc, Lombriser:2009xg}.
 
In order to recover the constraint relation \eqref{CSZ5} in the toy model, we need to give a different interpretation of these parameters in the sDGP model,
\begin{align}
\tO_{\L} &=    \Omega_{\Ld}  ,\qquad 
\tO_{D} =  \Omega_\DGP(t) -\O_\Ld,\qquad
\tO_{B}  \equiv   \frac{\Omega_M}{a(t)^3}= \frac{H(t)^2}{H_0^2} - \Omega_\DGP(t)\, , \label{DeltaB}\\
{\tw}_{D} &=-1-\frac{1}{3 H(t)} \frac{\dot  \Omega_\DGP(t)}{  \Omega_\DGP(t) -\O_\Ld }, \qquad~~~
 \Omega_\DGP(t)\equiv \frac{\rho_{\DGP}}{\rho_c} = \sqrt{\Omega_{\Ld}} \frac{H(t) }{H_0 }\, .\label{DeltaD}
\end{align}
In particular, taking the derivative of \eqref{DGPF1} and eliminating $\Omega_M$ with \eqref{DGPF1}  again will lead to the identical relation of $\dot{H}(t)$, as well as ${\tw}_{D}$ from \eqref{DeltaD},
\begin{align}
 {\dot{H}(t)} &=  - {3 H(t)^2} \frac{\frac{H(t)}{H_0}-\sqrt{\O_{\Ld}}}{\frac{2 H(t)}{H_0}-\sqrt{\O_{\Ld}} }, \label{dotDGP} \\
{\tw}_{D} &= - \frac{\frac{H(t)}{H_0}-\sqrt{\O_{\Ld}}}{\frac{2 H(t)}{H_0}-\sqrt{\O_{\Ld}} }\, . \label{wDGP}
\end{align}
One can check that the general constraint relation \eqref{CSZ5} is satisfied after putting back the expressions \eqref{DeltaB} and \eqref{wDGP}, at any cosmological time $t$.

From \eqref{DGPF1} and \eqref{wDGP}, it is clear to see that only in the very late time universe that $H(t)\to H_0\sqrt{\O_\Ld}$, we can have ${\tw}_{D}(t)\to 0$, which is a bit different with the $\L$CDM model.
If setting $\Omega_M=\Omega_B $, equating the right hand side of \eqref{DGPF} with that in the late time $\L$CDM model \eqref{LCDMF} at $a(t_0)=1$, we arrive at  $\Omega_D^2 =\Omega_{\L}\( \Omega_B-\Omega_D\)$. It is quite different from our toy constraint relation \eqref{CSZ1}, because it requires $\O_B>\O_D$, which does not match with the observed parameters in \eqref{OmegaD}. Thus, if taking the sDGP model with Friedmann equation in \eqref{DGPF} and setting $\Omega_M=\Omega_B $, it can not recover our toy constraint relation \eqref{CSZ1}.

In the next section, we will give an alternative interpretation of the embedding scenario as the holographic FRW universe. In particular, we will turn on the parameter $I$, which is usually set to be zero in the the previous studies of the  DGP models \cite{Lue:2005ya}.
It will become clear that only if keeping this parameter $I$ in the embedding function \eqref{Gat}, 
we can recover the constraint relation \eqref{CSZ1} in our toy model.

\subsection{Holographic Scenario: Holographic FRW Universe}
\label{sec33}

In the previous subsection, we have studied the dynamics of a FRW hypersurface, which is embedded into the higher dimensional flat spacetime. 
The physical picture is related to the traditional braneworld models \cite{Maartens:2003tw, ArkaniHamed:1998rs, ArkaniHamed:1998nn, Randall:1999ee, Randall:1999vf, Shiromizu:1999wj}, or the blackfold approaches with higher dimensional embedding \cite{Emparan:2009cs, Emparan:2009at, Emparan:2009vd, Carter:2000wv, Armas:2012ac, Armas:2012jg, Armas:2014bia}.
In this subsection, we will give a new physical interpretation of the FRW hypersurface in a flat bulk with the embedding metric \eqref{Gaussian}. It can be reduced to our toy model with a de-Sitter hypersurface in the flat bulk.
From the viewpoint of the cutoff holography in the flat spacetime \cite{Bredberg:2010ky, Bredberg:2011jq},
we can drop the manifold $\M_-$ in the flat bulk, such that the hypersurface $\p\M$ at $\w=0$ plays the role of   the holographic boundary of the manifold $\M_+$.  
Or from the viewpoint of membrane paradigm \cite{Parikh:1997ma}, the manifold $\M_+$ can be replaced by the quasi-local Brown-York stress-energy tensor on the hypersurface $\p\M$.
It is also equivalent to set  $\T_{\mu\nu}^{+} =0$ in the junction condition \eqref{junction1}, and the Einstein field equation becomes
\begin{align}
\frac{1}{\kappa_4} G_{\mu\nu}&=T_{\mu\nu}^{M} + \T_{\mu\nu}^{{\H}},
\end{align}
where the Brown-York stress-energy tensor  on $\p\M$ is
\begin{align}
 \T_{\mu\nu}^{\H}&=-\frac{2}{\sqrt{-g}} \frac{ \delta({\mS}_5)}{\delta g^{\mu\nu}}= \T_{\mu\nu}^{-} =\frac{1}{\kappa_5}\( {{\K}^-_{\mu\nu}} -{\K}^- g_{\mu\nu}\).
\end{align}
Again the Hamiltonian constraint equation  \eqref{Constraint0}  is automatically satisfied
\begin{align}
\frac{\T_{\H}^2}{3}-\T^{\H}_{\mu\nu}\T_{\H}^{\mu\nu} =
- \frac{ \kappa_4}{( \kappa_{5})^2} \big[ T_{\mM} + \T_{\H} \big].
\label{ConstraintH}
\end{align}

We will choose the negative branch in \eqref{Gat}, such that the energy density and pressure in $\T^{\H}_{\mu\nu}$ are given by
\begin{align}
(\kappa_5 c^2) \rho_{\H}  &= - \frac{3 \partial_\w\at(\w,t)}{\at(\w,t)}\Big|_{\w=0}=  3\Big[\frac{H(t)^2}{c^2} + \frac{\h}{a(t)^4}\Big]^{1/2} , \label{rhoH} \\
(\kappa_5 ) p_{\H}&=\[\frac{2\partial_\w\at(\w,t)}{\at(\w,t)}+\frac{\partial_\w\dot\at(\w,t)}{\dot\at(\w,t)}\]\Big|_{\w=0}
=-  \Big[ \frac{ \dot{H}(t)+3 H(t)^2 }{c^2} + \frac{\h}{a(t)^4} \Big]  \Big/\Big[\frac{H(t)^2}{c^2} + \frac{\h}{a(t)^4}\Big]^{1/2} .\label{pH}
\end{align}
The modified Friedmann equation is,
\begin{align}\label{Friedmann2}
H(t)^2 &= \frac{\kappa_4 c^4 }{3} \[ \rho_{\mM} (t) + \rho_{\H}(t)\],\qquad H_0^2  = \frac{\kappa_4 c^4 }{3}  \rho_{c}\, .
\end{align}
And the energy conservation equation remains
\begin{align}\label{conservation2}
\dot\rho_{\i}(t)&=-3 H(t) \[\rho_\i(t)+ p_\i(t)/c^2\],\qquad \i=\mM,\, {\H}\,.
\end{align}
Again we use the same setting in \eqref{kappaD}, considering that $\rho_{c}  =  \frac{3}{\kappa_4}\frac{H_0^2}{c^4 }$,
we have
 \begin{align}\label{kappar1}
 \Omega_{\L}=\frac{\rho_{\L} }{\rho_c}=\frac{c^2}{L^2 H_0^2},\qquad
\rho_{\L}  =  \frac{3}{\kappa_4}\frac{1}{c^2 L^2},\qquad
L=\frac{\kappa_5}{\kappa_4} .
 \end{align}

Putting \eqref{rhoH} into \eqref{Friedmann2}, the modified Friedmann equation is summarized as
\begin{align} \label{FRWOM}
 \frac{H(t)^2}{H_0^2} = \frac{\Omega_M}{a(t)^3} + \Omega_{\L}^{1/2} \[\frac{H(t)^2}{H_0^2} + \frac{\Omega_{\h}}{a(t)^4}\]^{1/2} ,\qquad \Omega_{\h}\equiv \frac{\h  c^2}{H_0^2}.
\end{align} 
Or equivalently,
\begin{align} \label{FRWF}
\frac{H(t)^2}{H_0^2}   = \frac{\Omega_{\L}}{2} + \frac{\Omega_M}{a(t)^3} + \[ \frac{\Omega_{\L}^2 }{4}+ \frac{ \Omega_{\L} \Omega_M}{a(t)^3} + \frac{ \Omega_{\L} \Omega_{\h}}{a(t)^4}\]^{1/2}.
\end{align}
We named this Scenario as the holographic FRW(hFRW) model. Instead of using the $\Lambda$CDM parameterization in \eqref{LCDMF}, we has a different set of parameters in the hFRW model. 
Notice here that by setting $\Omega_M=\Omega_B+\Omega_D$ and $\Omega_{\h}=0$, we can recover the usual Friedmann equation  \eqref{DGPF} of the sDGP model.
While if setting $\Omega_M=\Omega_B$ and turning  the parameter $\Omega_I$, it can be shown that one  is able to recover our toy constraint relation \eqref{CSZ1}.

Firstly, we need to match these parameters in hFRW model with that in the constraint relation \eqref{CSZ5},
\begin{align}
\tO_{\L} &= \Omega_{\L}, \qquad
\tO_{D} =  \Omega_\H(t) -\O_\L, \qquad 
\tO_{B}\equiv \frac{\Omega_M}{a(t)^3}
= \frac{H(t)^2}{H_0^2} -\Omega_\H(t), \label{hFRWF1}\\
{\tw}_{D} & =-1-\frac{1}{3 H(t)} \frac{\dot \Omega_\H(t)}{ \Omega_\H(t)-\O_\L},\qquad\quad
\Omega_\H(t)\equiv \frac{\rho_{\H}}{\rho_c}
= \Omega_{\L}^{1/2} \Big[ \frac{H(t)^2}{H_0^2} + \frac{\Omega_{\h}}{a(t)^4} \Big]^{1/2}\, .\label{hFRWF2}
\end{align}
In particular, taking the derivative of \eqref{FRWOM} and eliminating $\Omega_M$ with \eqref{FRWOM}  again will lead to the identical relation of $\dot{H}(t)$, as well as ${\tw}_{D}(t)$ from \eqref{hFRWF2},
\begin{align}
 {\dot{H}(t)} &=  - {3 H(t)^2}  \frac{\[ \sqrt{\frac{ H(t)^2}{H_0^2}+\frac{\O_I}{a(t)^4}} -\sqrt{\O_{\L}}\]
  - \frac{1}{3}\frac{\O_I}{a(t)^4}\big/\frac{H(t)^2}{H_0^2} }
 {2\sqrt{\frac{ H(t)^2}{H_0^2}+\frac{\O_I}{a(t)^4}}- \sqrt{\O_{\L}} }\, , \label{dotFRW} \\
{\tw}_{D}&= - \frac{\[\sqrt{\frac{ H(t)^2}{H_0^2}+\frac{\O_I}{a(t)^4}} -\sqrt{\O_{\L}}\]   - \frac{1}{3} \frac{\O_I}{a(t)^4}\big/\[\sqrt{\frac{ H(t)^2}{H_0^2}+\frac{\O_I}{a(t)^4}} -\sqrt{\O_{\L}}\]}{ 2\sqrt{\frac{ H(t)^2}{H_0^2}+\frac{\O_I}{a(t)^4}}- \sqrt{\O_{\L}}  }\,  . \label{wFRW}
\end{align}
One can check that the general constraint relation \eqref{CSZ5} is satisfied automatically
after plugging in above quantities \eqref{hFRWF1} and \eqref{wFRW}, at any cosmological time $t$. 


Now let us compare it with the late time evolution of $\Lambda$CDM model with Friedmann equation  \eqref{LCDMF}.
If only setting $\Omega_M=\Omega_B $, and equalizing the right hand side of \eqref{FRWF} and \eqref{LCDMF} at $a(t_0)=1$, we arrive at
\begin{align}\label{OmegaI0}
 \Omega_D^2 = \Omega_{\L} \Omega_{\h} -  \Omega_{\L}\( \Omega_D-\Omega_B\).
\end{align}
Thus, once taking
\begin{align} \label{OmegaI}
\Omega_{\h}=\frac{3}{2}\(\Omega_D-\Omega_B\),  
\end{align}
we can recover our toy constraint relation in \eqref{CSZ1}.
Plugging the  $\Lambda$CDM parameterization
\eqref{LCDMF} into the effective energy density \eqref{rhoH} and pressure \eqref{pH} from the Brown-York stress-energy tensor, and considering \eqref{OmegaI0}, we have
\begin{align}
\rho_\H\simeq\rho_c(\Omega_\L+\Omega_D),\qquad p_\H\simeq- (\rho_c c^2)\Omega_\L.
\end{align}
Thus, it is consistent with the ansatz in our toy model \eqref{BYDM} with $\rho_D=\rho_c\Omega_D$ and $p_{D}=0$, as well as $\rho_\Lambda=\rho_c\Omega_\L=- \frac{p_\Lambda}{c^2}$.

Finally, we summarize the normalized Hubble parameters $ {H(z)}/{H_0}$ in terms of the redshift $z$ in various models.
The redshift $z$ is related to the scale factor via $a(t)/a(t_0)=1/(1+z)$.
Considering \eqref{LCDMF}\eqref{DGPF}\eqref{FRWF} and setting $a(t_0)=1$, we have
\begin{align}
\Lambda\text{CDM}: ~
\frac{H(z) }{H_0} &=  \sqrt{ \Omega_\Lambda +   {\Omega_M}{(1+z)^3} }, 
\label{HzLCDM}\\\
\text{sDGP}: ~
\frac{H(z)}{H_0}  & = \sqrt{\frac{\Omega_{\Ld}}{2}+ {\Omega_M}{(1+z)^3}
+  \frac{\Omega_{\ell}}{2} \Big[ 1 +  \frac{4 \Omega_M}{\Omega_{\ell}} {(1+z)^3}\Big]^{1/2}},
\label{HzsDGP}\\
\text{hFRW}: ~ 
\frac{H(z) }{H_0} &  =\sqrt{  \frac{\Omega_{\L}}{2} + {\Omega_B}{(1+z)^3} 
+\frac{\Omega_{\Lambda}}{2} \Big[ 1+  
\frac{4 {\Omega_B}}{ \Omega_{\Lambda} } {(1+z)^3} + 
\frac{4 {\Omega_I}}{ \Omega_{\Lambda} } {(1+z)^4} \Big]^{1/2}
}.
\label{HzhFRW}
\end{align}
Taking \eqref{wDGP} and \eqref{wFRW}, the associated state equations ${\tw}_{D}(z)$ for various models are
\begin{align}
\Lambda \text{CDM}: ~ 
{\tw}_{D}(z)&= 0 \, , 
\label{wDLCDM}\\
\text{sDGP}:~ 
{\tw}_{D}(z)& =  - \frac{\frac{H(z)}{H_0}-\sqrt{\O_{\Ld}}}{\frac{2 H(z)}{H_0}-\sqrt{\O_{\Ld}} }\, ,
\label{wDsDGP}\\
\text{hFRW}: ~ 
{\tw}_{D}(z)&= - \frac{\[\sqrt{\frac{ H(z)^2}{H_0^2}+ {\Omega_I}{(1+z)^4}} -\sqrt{\Omega_{\Lambda}}\]   - \frac{1}{3}  {\O_I}{(1+z)^4}\Big/
\[\sqrt{\frac{ H(z)^2}{H_0^2}+ {\Omega_I}{(1+z)^4}} -\sqrt{\Omega_{\Lambda}}\]}{ 2\sqrt{\frac{ H(z)^2}{H_0^2}+ {\Omega_I}{(1+z)^4}}- \sqrt{\O_{\L}}  }\, .
\label{wDhFRW} 
\end{align}
Again in order to make the presentation simpler, we here neglected the contribution of radiation $\Omega_R$ and spatial curvature $\Omega_K$, which can be easily included in the equations above. Here including $\Omega_{\h}$ in \eqref{wDhFRW} turns the value of ${\tw}_{D}(z)$ from negative to positive in the late time universe, and thus effectively contributes to the emergent dark matter.
 
\begin{figure}[h]
\centering
\includegraphics[scale=0.55]{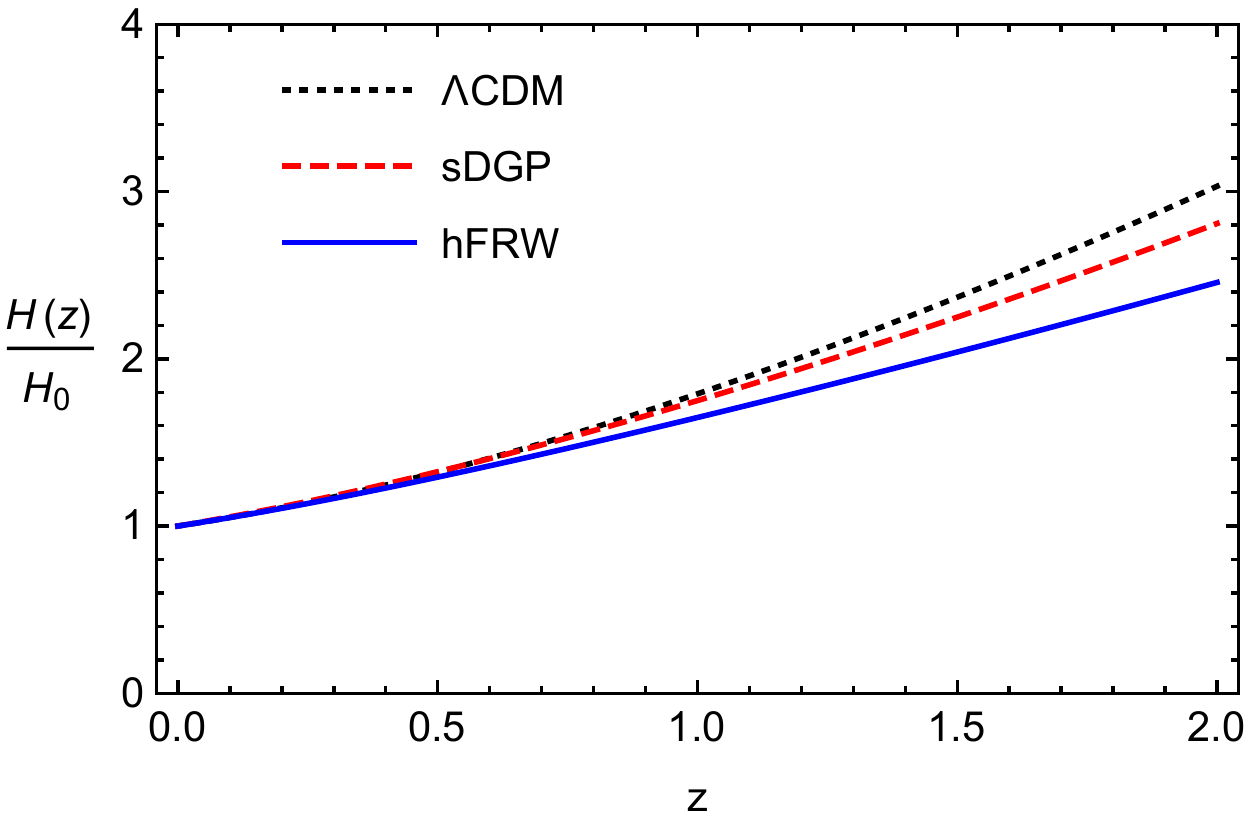}\qquad
\includegraphics[scale=0.55]{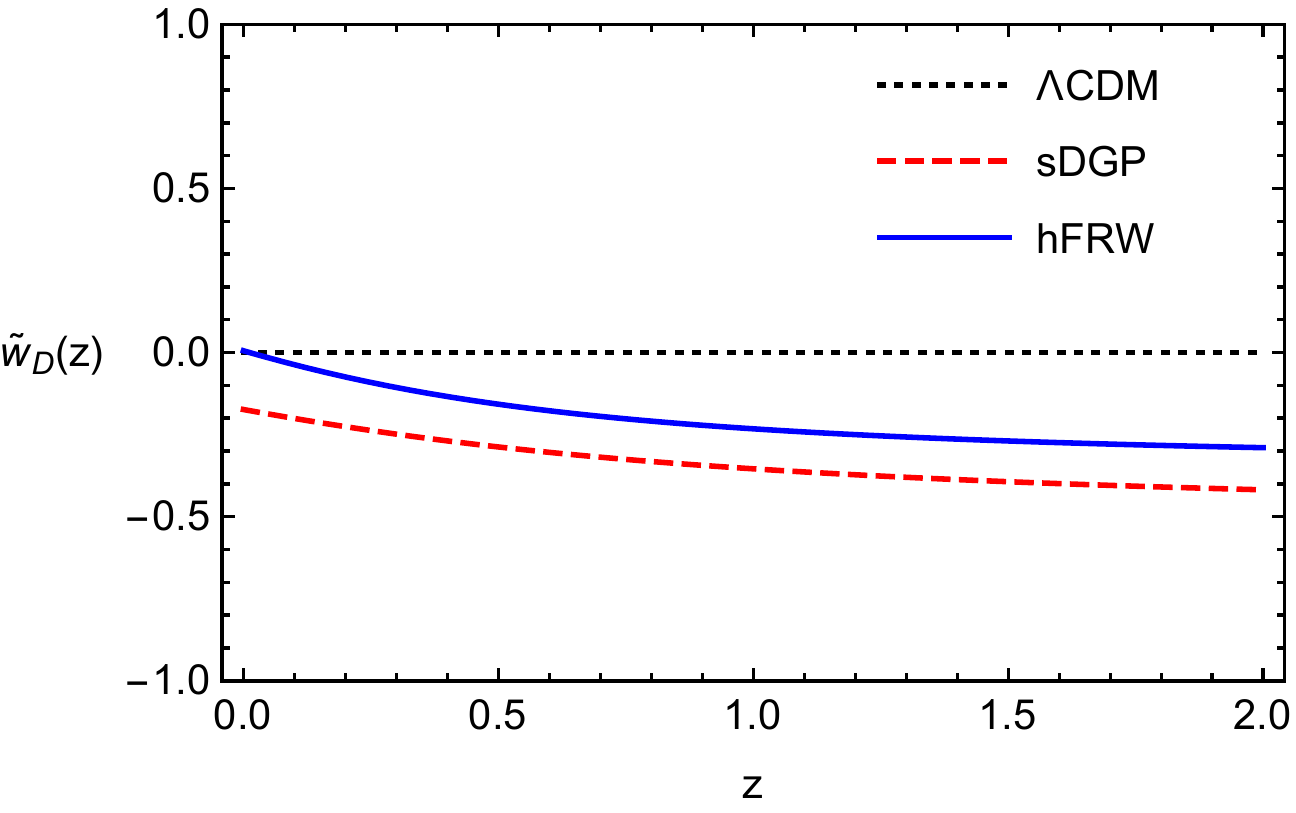}
\caption{Left: The reduced Hubble parameters $H(z)/H_0$ in terms of the redshift $z$ in various models.
Right: The evolution of state equations ${\tw}_{D}(z)$ in terms of the redshift $z$ in various models.
$\L$CDM: The plotting functions are in \eqref{HzLCDM} and \eqref{wDhFRW}, with the parameters in \eqref{OmegaD};
sDGP: The plotting functions are in \eqref{HzsDGP} and \eqref{wDsDGP}, with the fitting parameter $\O_M=0.21$ in \cite{Lue:2005ya};
hFRW: The plotting functions are in \eqref{HzhFRW} and \eqref{wDhFRW}, with a special choice of the parameters 
$\O_M=\O_B$, $\O_I=\frac{3}{2}(\O_D-\O_B)$, along with the values in \eqref{OmegaD}.
} \label{fig1}
\end{figure}

In Figure \ref{fig1}, we plot the reduced Hubble parameters $H(z)/H_0$ and the state equations ${\tw}_{D}(z)$ of the emergent dark matter in terms of the redshift $z$ in various models. 
The Friedmann equation of spatial flat $\L$CDM model is in \eqref{LCDMF}, with the  parameters in \eqref{OmegaD}. 
The Friedmann equation of sDGP model is in \eqref{DGPF}, with the fitting parameter $\O_M=0.21$  \cite{Lue:2002fe}.
The Friedmann equation of our hFRW model is in \eqref{FRWF}, 
with a special choice of the parameters $\O_M=\O_B, \O_I=\frac{3}{2}(\O_D-\O_B)$, along with the values in \eqref{OmegaD}.
More detailed studies of this non-zero $\Omega_{\h}$ and fitting parameters in the hFRW model will appear in our future work.





\section{Towards Holographic de Sitter Brane with Elasticity}
\label{Sec4}
 
In the above section \ref{Sec2}, inspired by the emergent gravity by Verlinde in \cite{Verlinde:2016toy}, we have proposed the emergent dark matter on the de-Sitter hypersurface in a flat bulk, which gives rise to the similar mechanism as in \cite{Verlinde:2016toy}.
In section \ref{Sec3}, we have generalized the holographic  de-Sitter scenario to the time evolution case with a FRW hypersurface in a flat bulk.
However, the above setups still lack of the elasticity in the Verlinde's emergent gravity \cite{Verlinde:2016toy}.
In the holographic models,  the elastic property can usually be realized in the blackfold approaches \cite{Emparan:2009cs, Emparan:2009at, Emparan:2009vd, Carter:2000wv, Armas:2012ac, Armas:2012jg, Armas:2014bia}, or by including the effective mass terms in the bulk of holographic models \cite{Alberte:2015isw, Alberte:2016xja, Alberte:2017cch}.
In this section, we will consider the embedding of a de Sitter hypersurface in the flat bulk with effective massive gravity terms, where the holographic elasticity is implemented.

On the other hand, in both section \ref{Sec2} and section \ref{Sec3}, we considered the uniform and isotropic metric at the cosmological scale. While in this section, aiming at a comparison with Verlinde's  derivation on the Tully-Fisher relation, we focus on the response at galactic scales of the de-Sitter brane. Instead of the uniform baryonic matter, we need to add the spherically symmetric baryonic matter.
We are also trying to reconcile the inconsistency in Verlinde's emergent gravity pointed out by \cite{Dai:2017qkz}. We present a consistent derivation of Tully-Fisher relation in the frame of the elastic model and try to resolve some issues in the original Verlinde's story.

\subsection{Holographic Stress Tensor and Verlinde's  Apparent Dark Matter}
\label{sec41}

 In order to embed the $d$-dimensional Verlinde's emergent gravity with elasticity into a higher dimensional bulk spacetime,
we sketch the more general total action as ${\mS}_{d+1}+{\mS}_{d}$, where
\begin{align}\label{Hmodel}
&{\mS}_{d+1}=\frac{1}{2 \kappa_{d+1}}\int_\M {\d}^{d+1} x \sqrt{-\tilde{g}}\[ \R_{d+1}-2 \L_{d+1} +\LM \]
+ \frac{1}{\kappa_{d+1}}\int_{\p\M} {\d}^{d}x \sqrt{-{g}}\mathcal{K}, \\
&{\mS}_{d}= \frac{1}{2 \kappa_d}\int_{\p\M} {\d}^dx \sqrt{-{g}} \(R_d -2\L_d\)+  \int_{\p\M} {\d}^{d}x \sqrt{-{g}} {\cal{L}}_{\mM} .
\end{align}
In the bulk manifold $\M$ with metric $\tilde{g}_{AB}$, the Lagrangian density $\LM$ represents the effective term which can provide the holographic elasticity. On the boundary $\p\M$ with induced metric ${g}_{\mu\nu}$, the trace of the extrinsic curvature is $\mathcal {K}$.
Like in our toy model in section \ref{Sec2}, we can set $\Lambda_{d+1}=0$ in the bulk, and study the holographic elastic response of the screen after putting in the baryonic matter.
One may also add $\Lambda_d$ in the boundary action ${\mS}_{d}$, which can contribute to the total cosmological constant on the boundary theory, or the tension of the boundary brane.

The viscous or elastic response of the boundary theory in the holographic description is encoded in the transverse-traceless tensor mode of the metric perturbations in the bulk. In the toy model in section \ref{Sec2}, we embed the $d$-dimensional de Sitter spacetime as the hypersurface in the $(d+1)$-dimensional flat bulk in Einstein gravity. The usual holographic solid model is on the flat boundary in AdS$_{d+1}$ spacetime in massive gravity  \cite{Alberte:2015isw, Alberte:2016xja, Alberte:2017cch}. However, one may embed the dS-sliced coordinates into the AdS$_{d+1}$ spacetime and obtain dS$_d$ as the boundary hypersurface. It is foreseeable that the elastic solid model can be generalized into the case with the de Sitter boundary with the tension term \cite{Marolf:2010tg, Buchel:2002wf, Alishahiha:2004md, Li:2011bt, Chu:2016uwi, Cai:2016lqa, Charmousis:2017rof, Gubser:1999vj, Savonije:2001nd, Kim:2001tb, Shiromizu:2001ve, Kanno:2002iaa, Apostolopoulos:2008ru, Kanno:2011hs}.
Although the detailed construction is still left to be done and we leave it to a future work, for now, we assume the above action can capture the feature of the elastic theory. In any case, the discussion for the rest of this section is actually independent of the holographic construction and it only uses the theory that describes the elastic solid with different modulus values in a de Sitter background.

For the $d$-dimensional de Sitter hypersurface embedded in the higher dimensional flat bulk spacetime, 
instead of the expanding coordinate in \eqref{dS0},
we now consider the static coordinate patch described by the metric
 \begin{align}\label{dSfr}
{\d}s_{d}^2 &={g}_{\mu\nu} {\d}x^\mu{\d}x^\nu  =-f(r){\d} t^2 +\hat{h}_{ij} {\d}x^i{\d}x^j ,\quad
f(r) = 1-\frac{r^2}{L^2},\\
{\d}s_{d-1}^2 &=\hat{h}_{ij} {\d}x^i{\d}x^j  = \frac{1}{f(r)} {\d} r^2 +r^2 {\d}{\Omega}_{d-2}^2,\quad
i, j= 1,2,... d-1.
 \end{align}
Here $\hat{h}_{ij}$ is the induced metric on the spacial slice.
One can also define the projection tensor $h^{\mu\nu}\equiv g^{\mu\nu} +u^\mu u^\nu$, along with the $d$ velocity $u^\mu$.
The Brown-York stress-energy tensor on the boundary of the bulk with the action \eqref{Hmodel}  is given by
\begin{align}\label{TmnSd}
\T^{\mu\nu}  &=  - \frac{2}{\sqrt{-{g}}} \frac{\delta{({\mS}_{d+1})} } {\delta {g}_{\mu\nu} }\equiv
\tilde{\rho}_D u^\mu u^\nu +\sg^{\mu\nu} ,
\end{align}
where $\tilde{\rho}_D$ is the effective holographic energy density induced from higher dimension, and we introduced the covariant stress tensor $\sg^{\mu\nu}$ which satisfies $\sg^{\mu\nu}u_\mu=0$.
The stress tensor $\sg_{ij}$ and strain tensor $\ve_{ij}$ are given by
\begin{align}\label{strain1}
\sg_{ij}\equiv {h}_i^\mu {h}_j^\nu\sg_{\mu\nu},\qquad
\ve_{ij} = {h}_i^\mu {h}_j^\nu\ve_{\mu\nu},\qquad  \ve_{\mu\nu}\equiv \nabla_{(\mu} \u_{\nu)},
\end{align}
and ${\u}_\mu$ is the shift vector associated with the deformation.

Now we take a detour to the Verlinde's derivation, where the displacement ${\u}_i$ is caused by the presence of the baryonic matter, and the metric solution in \eqref{dSfr} becomes
\begin{align}\label{newton}
f(r) = 1-\frac{r^2}{L^2}+2\Phi_B,\qquad \Phi_B\equiv -\frac{8\pi G_{d}}{(d-2)\Omega_{d-2}}\frac{M_B}{r^{d-3}}.
\end{align}
We consider the simple case that $M_B =\int ^{r_B}_0 \rho_D(r')A(r') dr'$ is the constant total mass of baryonic matter in the galaxy, with the characteristic scale $r_B$.
In the deep-MOND regime we are looking at in this section, $r_B < r\ll L$  \cite{Famaey:2011kh, Milgrom:1983ca},
such that $f(r) \approx 1+ 2\Phi_B$ will be taken in the following derivations \cite{Verlinde:2016toy}.

One issue in the Verlinde's derivation raised by \cite{Dai:2017qkz}, is that the displacement ${\u}_i=\frac{\Phi_B}{a_0}n_i$ is identified, with $a_0=c H_0$.
When choosing $d=4$ in \eqref{newton}, the baryonic matter induces a Newtonian potential $\Phi_B \sim -\frac{GM_B}{r}$, then the apparent dark matter surface density $\Sigma_D \sim \ve_{ij}$ scales as $1/r^2$ at the large $r$.
However, to produce the Tully-Fisher relation or flat rotational curves in galaxies, $\Sigma_D \sim \ve_{ij}$ has to scale as $1/r$ at the large $r$. Thus, this is the inconsistency in the Verlinde's original story  \cite{Dai:2017qkz}.
In the following, we try to circumvent this issue and see whether this assumption of the displacement can be abandoned. This displacement ${\u}_i=\frac{\Phi_B}{a_0}n_i$ is important  in \cite{Verlinde:2016toy}, where the ADM mass definition of the de Sitter spacetime is related to the strain tensor through
$M=\frac{1}{a_0}\oint _{{\mS}_\infty} \sigma_{ij}n_j d A_i$.
The problem with this argument is for a pure de Sitter space with the positive cosmological constant in Einstein gravity, it can not have the elastic property on its own. The elaborated derivation \cite{Verlinde:2016toy} avoids the facts that one needs to go beyond the theory of Einstein gravity to have the correct elastic dark matter.
One way out is to embed the de Sitter hypersurface in higher dimensional bulk, such that the elasticity will emerge from the holographic brane,
and the effective Einstein field equations will be modified.
In the following, we propose a way to resolve this issue by employing a model of holography with the bulk action \eqref{Hmodel} and reproduce the elastic dark matter response formula as the baryonic Tully-fisher relation in \cite{Verlinde:2016toy}.

\subsection{Emergent Tully-Fisher Relation from Holographic Elasticity}
\label{sec42}

Considering the fact that $\delta \ve_{ij}=-\delta \hat{h}_{ij} /2 $ in \cite{Alberte:2015isw, Alberte:2016xja, Alberte:2017cch}
and \eqref{TmnSd},
the spacial components are
 \begin{align}
 \sg_{ij}  &=\frac{1}{\sqrt{-g}} \frac{\delta(\sqrt{-g} \F ) }{\delta \ve^{ij}}
 =\frac{-2}{\sqrt{-g}} \frac{\delta( \mS_{d+1} ) }{\delta\hat{h}^{ij}}
=   2 \mu \ve_{ij} +  \lambda \delta_{ij}  (\ve^k_{~k}) , \label{Ela1}
\end{align}
with shear elastic modulus $\mu$ and bulk modulus $\lambda+2\mu/(d-1)$. 
By tuning the effective Lagrangian $\LM$ in the bulk \eqref{Hmodel}, we expect the designed values of $\mu$ coming out of the holographic engineering with only the traceless stress tensor \cite{Alberte:2015isw, Alberte:2016xja, Alberte:2017cch}, such that the bulk modulus vanishes $\lambda+2\mu/(d-1)=0$.
We can define the traceless part $ \ve'_{ij}$ as below,
\begin{align}\label{shear}
\sg_{ij}&=2 \mu \, \ve'_{ij},\qquad
 \ve'_{ij}  = \ve_{ij}- \frac{1}{d-1}  \delta_{ij} (\ve^k_{~k}) .
\end{align}
This describes only the shear deformation, without changing the volume of the body.  The  stress tensor $\sg_{ij}$ and strain tensor $\ve_{ij}$ then only have the traceless part $\ve'_{ij}$. 
We can diagonalize the elastic strain tensor $\ve_{ij}$ and stress tensor $\sg_{ij}$ simultaneously, since they are symmetric and linearly related. Their eigenvalues are called the principal strain and stress values.We define $\ve(r)$ as the largest eigenvalue of the traceless part of the strain tensor,
\begin{align}
\ve (r) &\equiv\ve'_{ij}  n^i n^j .
\end{align}
We adopt the volume formula of the entropy change $\Delta S(r)$ of the de Sitter spacetime by the total baryonic matter within a radius $r$ in \cite{Verlinde:2016toy},
\begin{align}
\Delta S(r)= - \frac{2 \pi \, M_B    r}{\hbar},\qquad T_{dS}= \frac{\hbar a_0}{2\pi },
\end{align}
where  $T_{dS}$ is the  Gibbons-Hawking temperature of de Sitter spacetime.
Notice that this relation is somewhat ad-hoc here since it assumes that holographic elastic de Sitter brane has the same entropy formula as the pure de Sitter space. We can put forth a naive argument that nevertheless, it can still have $\Delta S(r) \propto  M_B r$.
The change of the free energy then follows the volume law of the thermodynamics of the ``de Sitter medium'' with \begin{align}
&\Delta F (r)= - T_{dS} \Delta S(r) =a_0   M_B  r .\label{df1}
\end{align}

Now slightly different from assuming the displacement  ${\u}_i$ in Verlinde's \cite{Verlinde:2016toy},
 we start from the variation of the holographic free energy density $\cal F$ in \eqref{Ela1}.
Similarly, we will only consider the leading order contribution in terms of $M_B$ with a fixed background metric in \eqref{newton},
where the effects of $r^2/L^2$ will also be neglected.
If we only consider the shear modes in \eqref{shear} and do not consider the variation of background metric, the free energy density formula becomes
\begin{align}
\delta{(\sqrt{-g} \F)} &=\sqrt{-g}  \sg_{ij} \delta \ve^{ij} =  \mu \,\delta ( \sqrt{-g}  \ve'_{ij}\ve'^{ij}).
\end{align}
The change of total free energy $F=\int_{{\mathcal{V}}_{d-1}} \! {\F}$ within a radius $r$ is approximately
\begin{align}
&\Delta F (r) = \mu\int_0^r{\d}\rp\! A(\rp)\big( {\ve'}_{ij} {\ve'}^{ij}  \big)\gtrsim    \mu \frac{d-1}{d-2} \int_0^r {\d}\rp\! A(\rp) \ve(\rp)^2. \label{df2}
\end{align}
Here  the area is $A(r)=\Omega_{d-2}r^{d-2}$. We will take the last equal sign to be approximately true by defining $\ve(r)$ as the largest eigenvalue of the traceless part of $\ve_{ij}$ and assuming the other principal strains are equal and summed to $-\ve$. This step is similar to the Verlinde's derivation.
After differentiating both equations \eqref{df1} and \eqref{df2} with respect to $r$, we arrive at
\begin{align}\label{elastic}
\mu \frac{d-1}{d-2}  A(r) \ve(r)^2 =a_0  M_B \,.
\end{align}
One notices here that if we identify $M_D(r) =\int ^r_0 \rho_D(r')A(r') dr'$ as the total apparent dark matter mass enclosed inside radius $r$,
and assume the surface density of the apparent dark matter $\Sigma_D$ and baryonic matter $\Sigma_B$ as
\begin{align}
\Sigma_D(r) \equiv \frac{M_D(r)}{A(r)}= \frac{\mu}{a_0} \ve(r),\qquad  \Sigma_B(r) \equiv \frac{M_B}{A(r)},
\end{align}
then from \eqref{elastic} we can arrive at the response relation as below,
\begin{align}\label{surface}
\Sigma_D(r)^2 = \frac{\mu}{a_0} \frac{d-2}{d-1} \Sigma_B(r),\qquad \mu = \frac{ a_0^2}{16 \pi G}.
\end{align}
The value of shear elastic modulus $\mu$ has been chosen to be the same as in Verlinde's \cite{Verlinde:2016toy},
although we take a different ansatz for the pre-factor in $\Sigma_D(r)$.

When $d=4$, and considering $ g_{D}(r) = {GM_{D}(r)}/r^2$, $ g_B(r) = {GM_B }/r^2$, the equation \eqref{surface} leads to the same relation in Verlinde's \cite{Verlinde:2016toy},
\begin{align}
g_D(r)^2= \frac{a_0}{6} g_B(r).
\end{align}
 In the deep-MOND regime that $g_{D}(r)\gg g_{B}(r)$, the above conclusion leads to the baryonic Tully-Fisher relation that,
\begin{align}
v_f^4 =  \frac{a_0}{6}GM_{B} , \qquad g_{D}(r) =\frac{v_f^2 }{r} ,
\end{align}
where $v_f$ is the asymptotic velocity of the flattened galaxy rotation curve.
 
Notice here that we haven't explicitly constructed the theory with baryonic mass $M_B$ on top of the elastic background. To do so requires more ingredients in the theory and it can become complicated, for such examples, see e.g. \cite{Buchel:2017qwd, Buchel:2017lhu, Buchel:2016cbj}.
We may compare the dark matter density here as in the toy model in section \ref{Sec2}.
The strain tensor $\epsilon_{\mu\nu}$ in \eqref{strain1} is related to the extrinsic curvature $\K_{\mu\nu}$.
The stress tensor $\sg_{\mu\nu}$ in \eqref{shear} contributes to the total Brown-York stress-energy tensor $\T_{\mu\nu}$.
For the toy model in section\ref{Sec2}, we employ the hypersurface displacement in extrinsic curvature as the elastic displacement tensor, whereas in the holographic model in this section, the extra fields with $\LM$ in the bulk is introduced. The boundary term in action \eqref{Hmodel} is also related to the Brown-York stress-energy tensor in the modified Einstein equations \eqref{Induced1} in the toy model.
This leads to the questions in the previous section that whether we can realize this response in a model with braneworld, although the answer is not clear at this moment.

\section{Discussion and Conclusion}  \label{SecOther}

In this final section, we will further discuss the difference and connections between our approach and some well-studied scenarios, especially the braneworld and the holographic universe.
Following the Verlinde's model for the apparent dark matter, there seem to be three crucial conditions for his construction so far. First, there is the background entropy, which distributes evenly in the volume. Second, the positive cosmological constant, provides the thermal bath with the Gibbons-Hawking temperature $T_{dS}=\hbar a_0/2\pi$. The last, the apparent dark matter is only the response to the presence of the baryonic matter.
The braneworld scenario may offer something similar to above-mentioned conditions and becomes a natural playground for the Verlinde's emergent gravity. The elastic medium full of entropy can be explained from higher dimensions though additional brane dynamics, by treating our (3+1) dimensional spacetime as the boundary of the bulk spacetime. Cosmological constant and the standard model can be easily implemented with braneworld in the literature \cite{Frieman:2008sn, Kaplan:2011vz}.
Most interestingly, the branes with tensions and dynamics, may react to the matter fields we put in, with extra terms introduced. Especially the extrinsic curvature, a concept valid only from higher dimensional spacetime, may describe the ``elastic response" nature of the apparent dark matter from the Verlinde's theory.

We comment on the possible constraints from gravitational wave observations.
Recently it is argued that two relativistic models of modified Newtonian dynamics seem inconsistent with observations \cite{Chesler:2017khz}. Modified gravity models are constrained from two aspects.
One is the constraint of the energy loss rate from ultra high energy cosmic rays, which indicates that gravitational waves should propagate at the speed of light.
The other is the observed gravitational waveforms from LIGO, which are consistent with Einstein's gravity and suggest that the gravitational wave should satisfy linear equations of motion in the weak-field limit.
Although Verlinde suggested similar modifications of Newtonian dynamics as in MOND theory \cite{Verlinde:2016toy}, which emerges with a different underlining physical origin,
there are no covariant equations of motion for the gravitational waves. For our toy model in the previous sections,
the induced dark components can be viewed as dark energy and dark matter stress-energy tensor and they behave the same as the particle dark matter in the $\Lambda$CDM models at the leading order,
so it can pass the above-mentioned constraints \cite{Chesler:2017khz}.
To be more specific, the extra apparent dark sector as the extra term in Einstein equation fills the space as dark medium and interacts with propagating photons in it. In our induced gravity, gravitational field equations are modified as
\begin{equation}\label{Einstein1}
 {R}_{\mu\nu}-\frac{1}{2}g_{\mu\nu} {R} = \kappa_4 \big[ {T}_{\mu\nu}+{\T}_{\mu\nu}\big].
\end{equation}
The Bianchi identity leads to
$0\equiv \nabla^\mu G_{\mu\nu}= \kappa_4 \nabla^\mu  {T}_{\mu\nu}+ \kappa_4 \nabla^\mu {\T}_{\mu\nu}  $.
If we did not put additional sources in the bulk, the Brown-York stress-energy tensor  \eqref{BY1} itself is conserved $\nabla^\mu {\T}_{\mu\nu}=0$.
Thus it is similar to the effects of real dark matter, and it does not conflict with the observations from LIGO so far \cite{Green:2017qcv}.

Let us further compare our toy model to the other well studied braneworld models \cite{Maartens:2003tw, ArkaniHamed:1998rs, ArkaniHamed:1998nn, Randall:1999ee, Randall:1999vf, Shiromizu:1999wj}
other than DGP.  Except for the constraint equations,
we also have the dynamical Einstein equations in higher dimensions.
For example, in BDL (Binetruy-Deffayet-Langlois) model \cite{Binetruy:1999ut},
 the FRW metric is also embedded into one higher dimensional flat spacetime.
After including the baryonic matter $T_{\mu\nu}$ on the brane, in principle we can also define the effective stress-energy tensor from the braneworld model
\begin{align}
&\T^\B_{\mu\nu}  \equiv \W_{\mu\nu} + (\K g_{\mu\sigma}- \K_{\mu\sigma})\K^\sigma_{~\nu}- \frac{1}{2}\(\K^2-\K_{\rho\sigma}\K^{\rho\sigma}\)g_{\mu\nu}+...\, ,
 \end{align}
which includes the description of  the hypersurface evolution, and $\W_{\mu\nu}$ is associated with the bulk Weyl tensor. It might be interesting to derive the Tully-Fisher relation based on this formula.
One thing we notice is that in most of the braneworld models \cite{Maartens:2010ar}, the gravity leaks to the extra dimension at large scale, so the gravitational force scales like $1/r^{d-2}$, comparing to the Newtonian force $1/r^2$. To have the observed flat galaxy rotational curves, the gravitational force has to scale like $1/r$. Naively this does not work. 
While one may try to extend the DBI action of the branes with more fields to capture the elastic behaviors of the dark matter response theory, see for example \cite{Bielleman:2016grv, Carone:2016tup}. One not only needs to add tension term to the D-brane action as the cosmological constant, the extra scalar/vector fields and its response with baryonic matter are also needed, see for example \cite{Hossenfelder:2017eoh, Dai:2017guq}.
The extra dimensions in the braneworld setups may also have some extra effects on the gravitational waves production and propagation. If we indeed take a higher dimensional point of view, we expect one extra breezing mode on top of two polarized propagating modes \cite{Kobayashi:2005dd, Andriot:2017oaz}. The extra breezing mode is constrained by the current experiments. There are also multiple massive Kaluza-Klein gravitational modes associated with the extra dimensions. Although those massive modes decay fast and may not reach the gravitational waves detector, they are constrained as well by the gravitational waves signal templates from the binary black hole signals.
Our model does not necessarily have observable effects from extra dimensions. It is still quite interesting to ask whether we can probe extra dimensions, although there is no evidence from experiments so far.  One recent study may come from gravitational waves physics in \cite{Andriot:2017oaz}, in addition to the long searching constraints from colliders and precision measurements of gravity.

In the traditional holographic theories for our $3{+}1$ dimensional universe \cite{tHooft:1993dmi, Susskind:1994vu, Strominger:2001pn, McFadden:2009fg}, gravity in the bulk is encoded into the field theory on the boundary.
For the models which consider the universe as the boundary of  $4{+}1$  dimensional AdS \cite{Gubser:1999vj, Savonije:2001nd, Kim:2001tb, Shiromizu:2001ve, Kanno:2002iaa, Apostolopoulos:2008ru, Kanno:2011hs}, there is an effective contribution from the holographic stress-energy tensor, which can be identified as the stress-energy tensor of dark energy and/or dark matter.
In our construction, the holographic screen is embedded into one higher dimensional flat spacetime, which is inspired by the holographic hydrodynamics in Rindler spacetime~\cite{Bredberg:2010ky, Bredberg:2011jq, Compere:2011dx, Compere:2012mt, Eling:2012ni,Lysov:2011xx, Cai:2013uye, Cai:2014ywa, Khimphun:2017bac}.
It is named as Rindler fluid, which is described by the Brown-York stress-energy tensor on the accelerating cutoff hypersurface in a flat bulk. The dynamics is governed by the Hamiltonian and momentum constraint equation on the hypersurface. 
What is more, it will be interesting to see how the entanglement can happen in our toy model with the de Sitter boundary \cite{Maldacena:2012xp}. The entanglement between two cosmological horizons may have an impact on the gravity as suggested by Verlinde~\cite{Verlinde:2016toy}.
It is shown in \cite{Jensen:2013ora, Sonner:2013mba, Chernicoff:2013iga, Chen:2016xqz}, that the entangled pair in $3{+}1$ dimensions can be described by the wormhole in $4{+}1$ dimensional bulk spacetime. Thus, it is more clear to see the entanglement through the embeddings of wormholes into a higher dimensional bulk.

In summary, we construct a model for the dark components of our universe, where the dark sector originates from the induced stress-energy tensor of higher dimensional spacetime. In this holographic picture, there is only baryonic matter and radiation in the late time universe and dark matter is considered as the response of baryonic matter from the geometric effects.
In our approach, the toy model and the more developed hFRW model are partly borrowed from the Verlinde's emergent gravity in a subtle way. We choose to start from a holographic de Sitter screen in higher dimensional flat spacetime, with the known covariant relativistic formulas. The toy model produces one additional constraint of the late time universe components from $\Lambda$CDM parameterization. We then relate our toy model to the DGP braneworld with the new interpretation of the dark matter.
Moreover, we suggest a new holographic scenario with a set of parameters for the late time universe evolution. 
Although it has been pointed out that there are some inconsistencies in the Verlinde's emergent gravity \cite{Milgrom:2016huh,Dai:2017qkz},
the idea that considers the dark matter as the response of baryonic matter is still quite interesting \cite{Cadoni:2017evg}. In section \ref{Sec4} we fix an inconsistency of Verlinde's story and re-derived the Tully-Fisher relation.
In the future, it is interesting to relate our holographic model to the other well-motivated dark matter models (see for example \cite{Kamada:2016euw, Tulin:2013teo,Berezhiani:2015pia,  Cai:2015rns}), 
as well as the emergent cosmology from quantum entanglement or thermodynamical laws (see for example  \cite{VanRaamsdonk:2010pw, Ge:2017rak, Cai:2005ra, Cai:2006rs}). 

\section*{Acknowledgments}

We thank  Y.\, F.\, Cai, S.\, P.\, Kim, B.\, H.\, Lee, T.\, Liu, C.\, Park, M.\, Sasaki, J.\, Soda, H.\, Tye, S.\, J.\, Wang, M. Yamazaki
for helpful conversations,  as well as the anonymous referees for very helpful comments and suggestions.
R.\, G.\, Cai was supported by the National Natural Science Foundation of China (No.11690022, No.11375247, No.11435006, No.11647601), Strategic Priority Research Program of CAS (No.XDB23030100), Key Research Program of Frontier Sciences of CAS.
S.\, Sun was supported by MOST and NCTS in Taiwan.
Y.\, L.\, Zhang was supported by Young Scientist Training Program in APCTP, which is funded by the Ministry of Science, ICT \& Future Planning(MSIP), Gyeongsangbuk-do and Pohang City.

\end{document}